\documentclass[a4paper,12pt]{article}
\pdfoutput=1 % if your are submitting a pdflatex (i.e. if you have
\usepackage{jcappub} % for details on the use of the package, please
\usepackage{epstopdf}
\usepackage{hyperref}
\hypersetup{
    bookmarks=true,         % show bookmarks bar?
    unicode=false,          % non-Latin characters in Acrobat’s bookmarks
    pdftoolbar=true,        % show Acrobat’s toolbar?
    pdfmenubar=true,        % show Acrobat’s menu?
    pdffitwindow=false,     % window fit to page when opened
    pdfstartview={FitH},    % fits the width of the page to the window
    pdftitle={MHI},    % title
    pdfauthor={Author},     % author
    pdfsubject={Subject},   % subject of the document
    pdfcreator={Creator},   % creator of the document
    pdfproducer={Producer}, % producer of the document
    pdfkeywords={keyword1} {key2} {key3}, % list of keywords
    pdfnewwindow=true,      % links in new window
    colorlinks=true,       % false: boxed links; true: colored links
    linkcolor=red,          % color of internal links
    citecolor=cyan,        % color of links to bibliography
    filecolor=magenta,      % color of file links
    urlcolor=blue,           % color of external links
    linktocpage=true
}
\usepackage{subfigure}
\usepackage{color}
\usepackage{amsfonts}
\usepackage{mathrsfs}
\usepackage{epsfig}
\usepackage{amsmath,amssymb,amsthm,graphicx,latexsym}
\DeclareMathOperator{\sech}{sech}
\usepackage{ulem}

\newcommand{\be}{\begin{equation}}
\newcommand{\ee}{\end{equation}}
\newcommand{\bea}{\begin{eqnarray}}
\newcommand{\eea}{\end{eqnarray}}
\newcommand{\fig}{Fig.\ref}
\newcommand{\eq}{Eq.\eqref}

\def\bea{\begin{eqnarray}}
\def\eea{\end{eqnarray}}
\def\ba{\begin{array}}
\def\ea{\end{array}}

\def\beq{\begin{equation}}
\def\eeq{\end{equation}}

\title{%\boldmath \bf 
An Inflationary Equation of State }

\author[a]{Barun Kumar Pal}

\affiliation[a]{Department of Mathematics \\  Netaji Nagar College for Women, Kolkata 700 092, INDIA}

\emailAdd{terminatorbarun@gmail.com}

\abstract{ We have studied inflationary paradigm through an inflationary equation of state. With  a single parameter equation of state as a function of the scalar field responsible for accelerated expansion, we find an observationally viable model  satisfying all the constraints as laid down by the recent observations. The resulting model can efficiently cover a wide range of tensor-to-scalar ratio ranging from $r\sim\mathcal{O}(10^{-1})$ to $\mathcal{O}(10^{-6})$, other inflationary observables being consistent with the latest data. Nowadays ultimate eliminator between inflationary models is the tensor-to-scalar ratio, the model presented here is capable of keeping up with the future probes of tensor-to-scalar ratio at the same time having good agreement with other inflationary observables. }

\begin{document} 
\maketitle
\flushbottom

%%%%%%%%%%%%%%%%%%%%%%%%%%%%%%%%%%%%%%%%%%%%%%%%%%%%%%%%%%%%%%%%%
\section{Introduction} \label{sec1}
More than four decades has elapsed since inflationary paradigm \cite{starobinsky1978,starobinsky1979, guth1981,mukhanov1981,guth1982, starobinsky1982,ovrut1983, linde1983, linde1983b,steinhardt1984, lucchin1984} started its journey to resolve the puzzles of standard Big-Bang cosmology.  Though Big-Bang theory has been complemented successfully, inflation has stayed as a paradigm due the non-existence of a unique compelling model. Recent observations \cite{akrami2020planck, aghanim2020planck, ade2014planck, spergel2007} have ruled out many inflationary models but still allowing plenty. As a consequence  inflationary model building is still alive with added charm coming from availability  of late time finespun data. The present and futuristic Cosmic Microwave  Background Radiation (CMB) missions in the likes of BICEP2/Keck \cite{ade2018constraints}, CMB-S4 \cite{abazajian2016cmb}, LiteBird \cite{matsumura2014mission} are promising to detect tensor-to-scalar ratio $\mathcal{O}(10^{-4})$ which will further reduce the number of viable  inflationary models.  
 
In general inflationary paradigm is investigated by specifying inflaton potential which may have field theoretic motivation or may be of purely phenomenological origin. Recently inflationary scenario is examined from hydro-dynamical point of view \cite{mukhanov2013quantum} by suitably selecting the equation of state parameter (EoS from now on) during observable inflation as a function of e-foldings. This EoS formalism is capable of explaining the whole inflationary scenario independent of any model and the observable outcomes can be confronted with the recent data at ease.   

In this article we have studied single field inflationary paradigm with a phenomenological EoS which is a function of scalar field responsible for accelerated expansion employing Hamilton-Jacobi formulation.  The EoS we are interested herein is given by
\beq
1+\omega=\frac{2}{3}\sech^p\left(\phi{\rm M_P}^{-1}\right)
\eeq 
where $p$ is any dimensionless positive number. The  Hamilton-Jacobi framework provides a simple one-to-one correspondence between inflationary EoS and  Hubble parameter and the associated potential. We shall see that for any positive value of p, above EoS has inflationary solution consistent with recent observations. We shall also see that p is inversely proportional to logarithm of tensor-to-scalar ratio, which helps to determine a lower bound of the model parameter depending on present upper bound  $r<0.064$ \cite{akrami2020planck}. We have found that the particular solution for $p=2$ mimics $\alpha$-attractor class of inflationary models \cite{kallosh2013}. We also find that a single parameter inflationary  EoS is consistent of latest data coming from various probes. 

This article is organized  as follows, in Sections \ref{hj} and \ref{mukhanov} we have briefly reviewed the Hamilton Jacobi formalism and Mukhanov Parametrization of single field inflation respectively. In Section \ref{model} we have discussed outcomes of the model under consideration. In Section \ref{p=2} we have shown that for $p=2$ our model represents a particular case of $\alpha$-attractor inflationary models. Finally we conclude in Section \ref{conclusion}.

%%%%%%%%55555555555555555555555555555555555555555%%%%%%%%%%%%%%%%%%%%%%%%%%%%%%%%%%%%%%%%%%%%%%%%%%%%
\section{Hamilton Jacobi Formalism }\label{hj}
Within the framework of  Hamilton-Jacobi formulation of inflation,  Friedmann equations can be rewritten as  \cite{salopek1990, muslimov1990,liddle1994, kinney1997,lidsey1997, barunquasi,barun2018mutated, barunmhi, barunmhip}
\bea
\left[\rm H'(\phi)\right]^2 -\frac{3}{2\rm {\rm M_{P}}^2}
{\rm H}(\phi)^2&=&-\frac{1}{2\rm {\rm M_{P}}^4}V(\phi)\label{hamilton}\\
\dot{\phi}&=&-2\rm {\rm M_{P}}^2 H'(\phi)\label{phidot}
\eea
where prime and dot denote derivatives with respect to the scalar field $\phi$ and time respectively, and ${\rm M_{ P}}\equiv\frac{1}{\sqrt{8\pi G}}$ is the reduced Planck mass. 
We also have 
\beq\label{adot}
\frac{\ddot{a}}{a}=H^2(\phi)\left[1-\rm\epsilon_{_H}\right] 
\eeq
where $ \epsilon_{_{\rm H}}$ has been defined as
\beq\label{epsilon} \epsilon_{_{\rm H}}=2\rm {\rm M_{P}}^2\left(\frac{H'(\phi)}{H(\phi)}
\right)^2.
\eeq 
So inflation occurs when $\epsilon_{_{\rm H}}<1$ and ends exactly at  $\epsilon_{_{\rm H}}=1$. So requirement for the violation of strong energy condition is uniquely determined only by  $\epsilon_{_{\rm H}}<1$. In general these equations are very complicated to solve without invoking to the particular form of the Hubble parameter.

The amount of inflation is represented by e-foldings which is defined as
\beq\label{efol}
\text{N}\equiv\ln\frac{a_{\rm end}}{a}=\frac{1}{\rm M_P}\int_{\phi_{\rm  end}}^{\phi}\frac{1}{\sqrt{2\epsilon_{_{\rm H}}}}\ d\phi.
\eeq 
The e-folding has been defined in such a way that at the end of inflation $\text{N}=0$ and $\text{N}$ increases as we go back in time. In order to solve the big-bang puzzles and to comply with recent observations we need  $50-70$ e-foldings before the end of inflation. 
Another conventional parameter is 
\begin{equation}\label{eta}
\eta_{_{\rm H}}=2{\rm {\rm M^2_{P}}}~ \frac{\rm H''(\phi)}{\rm H(\phi)}. 
\end{equation}
It's worthwhile to mention here that $\epsilon_{_{\rm H}}$ and $\eta_{_{\rm H}}$ are not the usual  slow-roll parameters,  $\epsilon_{_{\rm H}}$ measures the relative contribution of the inflaton's kinetic energy to its total energy, whereas $\eta_{_{\rm H}}$ determines the ratio of  field’s acceleration relative to the friction acting on it due to the expansion of the universe \cite{lidsey1997}.
\iffalse
Though we have not included higher order slow-roll parameters in the present analysis, following parameters are widely used,
\bea
\zeta_{_H}^2(\phi)&\equiv& 4{\rm {\rm M_{P}}}^4\ \frac{H'(\phi)H'''(\phi)}{H^2(\phi)}\\
\sigma_H^3(\phi)&\equiv& 8{\rm {\rm M_{P}}}^6\ \frac{H'^2(\phi)H''''(\phi)}{H^3(\phi)}.
\eea \fi
Slow-Roll approximation applies when $\{\epsilon_{\rm H}, |\eta_{\rm H}|\ll 1\}$. Inflation goes on as long as  $\epsilon_{\rm H}<1$, even if slow-roll is broken. The breakdown of slow-roll approximation  drag the inflaton towards its potential minima and end of inflation happens quickly.

%%%%%%%%%%%%%%%%%%%%%%%%%%%%%%%%%%%%%%%%%%%%%%%%%%%%%%%%%%%%%%%%%%%%%%%%%%%%%%%%%%%%%%%%%%%%
\section{Mukhanov Parametrization}\label{mukhanov}
It is quite remarkable that the whole inflationary evolution can be  represented efficiently by suitably picking the EoS as described in Ref.\cite{mukhanov2013quantum}. In this parametrization single field inflation can be studied model independently. Within the framework of Hamilton-Jacobi formulation, inflationary EoS is exactly given by 
\beq
1+\omega=\frac{2}{3}\epsilon_{_H}.
\eeq
For accelerated expansion we need $\omega<-\tfrac{1}{3}$. A cosmological constant would give us $\omega=-1$, but then we will have eternal inflation which is observationally forbidden. So we need an equation of state which will provide sufficient amount of inflation along with graceful exit from inflation. Given a model i.e. potential involved or the \textit{Hubble} parameter, inflationary dynamics can be solved but with a specific EoS  underlying scenario may be investigated model independently. In Ref.\cite{mukhanov2013quantum} inflationary paradigm has been investigated with the following parametrization  of  EoS
\beq
1+\omega=\frac{\beta}{({\rm N}+1)^\delta}
\eeq 
where $\delta \ \mbox{and} \ \beta$ are positive constants, which captures a wide range of inflationary models with various observational predictions \cite{gariazzo2017primordial}.

Considering $\text{N}$ as new time variable it is possible to express all the inflationary observables in terms of the EoS \cite{mukhanov2005, garcia2014large}, up to the first order in slow-roll parameters,  
\bea
n_{_S}-1 &\simeq& -3\left(1+\omega\right) + \dfrac{d}{d\text{N}}\ln\left(1+\omega\right)\\
\alpha_{_S} &\simeq& 3\dfrac{d}{d\text{N}}\left(1+\omega\right)-\dfrac{d^2}{d\text{N}^2}\ln\left(1+\omega\right)\\
r&\simeq&24\left(1+\omega\right)\\
n_{_T}&\simeq&-3\left(1+\omega\right)\\
\alpha_{_T} &\simeq&3\dfrac{d}{d\text{N}}\left(1+\omega\right).
\eea
To confront with the recent observational data above quantities are evaluated at the time of horizon crossing i.e. when there are 50-70 e-foldings left before the end of inflation.

% So with the specific equation of state parameter one can get a very good estimate of the observable parameters. 

%%%%%%%%%%%%%%%%%%%%%%%%%%%%%%%%%%%%%%%%%%%%%%%%%%%%%%%%%%%%%%%%%%%%%%%%%%%%%%%%%%%%%%%%%%
\section{Equation of State}\label{model}
It's evident that a slowly varying EoS  will produce sufficient amount of inflation along with graceful exit. In Ref.\cite{mukhanov2013quantum} single field inflation scenario has been investigated with EoS as a direct function of e-folding, instead, in this work we have examined inflationary paradigm with EoS which is a function of the scalar field responsible for inflation having the following form
\beq\label{eos_def}
1+\omega=\frac{2}{3}\sech^p\left(\phi{\rm M_P}^{-1}\right)
\eeq 
where $p$ is a dimensionless positive number. We have imposed the positivity restriction on the parameter $p$ in order to get inflationary solutions. As $p<0$ would imply decelerated expansion which is clear from \eq{adot}. Apart from this positivity constraint the model  parameter $p$ is otherwise free. From \eq{eos_def} we see that $\omega<-\tfrac{1}{3}$ as long as $ |\phi|>0$. So for this class of inflationary models, accelerated expansion goes on as long as $|\phi|>0$ and ends exactly at $\phi=0$. In Ref.\cite{german2021new} a generalization of the $\alpha$-Attractor T models has been proposed. Where inflationary scenario has been investigated with the following potential
\beq
\rm V=V_0\left(1-\sech^p(\lambda\phi/{\rm M}_P)\right).
\eeq
In this article our approach is somewhat different and we shall see that only $p=2$ closely resembles $\alpha$-Attractor model for $\alpha=2/3$. 

Given the EoS it is now very simple to get the corresponding Hubble parameter which, within  the Hamilton-Jacobi formulation, has the following exact form
\beq
{\rm H}={\rm H_0} \ e^{\frac{1}{\sqrt{2} }\sinh (\phi{\rm M_P}^{-1} ) \, _2F_1\left[\frac{1}{2},\frac{1}{2}+\frac{p}{4};\frac{3}{2};-\sinh ^2(\phi{\rm M_P}^{-1} )\right]}\label{EOS_Hubble},
\eeq
where $ _2F_1$ is Hyper-geometric function.
\begin{figure}%[htb]
	\centerline{\includegraphics[width=16.cm, height=9.cm]{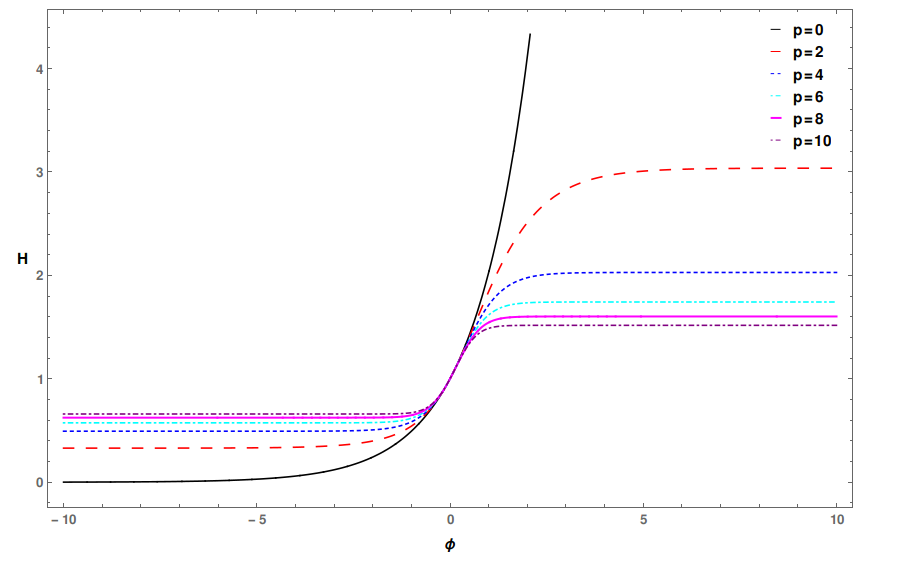}}
	\caption{\label{fig_hubble} The variation of Hubble parameter (in the units of ${\rm H}_0$) with the scalar field for six different values of the model parameter, $p\in [0,10]$.}
\end{figure}
In \fig{fig_hubble} we have shown the variation of inflationary Hubble parameter  with the scalar field for six different values of the parameter $p$. From the figure it is  clear that apart from $p=0$, the Hubble parameter varies slowly which essentially renders sufficient period of accelerated expansion. We have included   $p=0$ case in our plots for illustration purpose only.

Corresponding potential can be determined using \eq{hamilton}  which is given by
\beq
{\rm V}={\rm H_0}^2 {\rm M_P}^2 \left(3-\text{sech}^p(\phi{\rm M_P}^{-1} )\right) e^{\sqrt{2} \sinh (\phi{\rm M_P}^{-1})\, _2F_1\left(\frac{1}{2},\frac{1}{2}+\frac{p}{4};\frac{3}{2};-\sinh ^2(\phi{\rm M_P}^{-1})\right)}. 
\eeq
\begin{figure}%[htb]
	\centerline{\includegraphics[width=16.cm, height=9.cm]{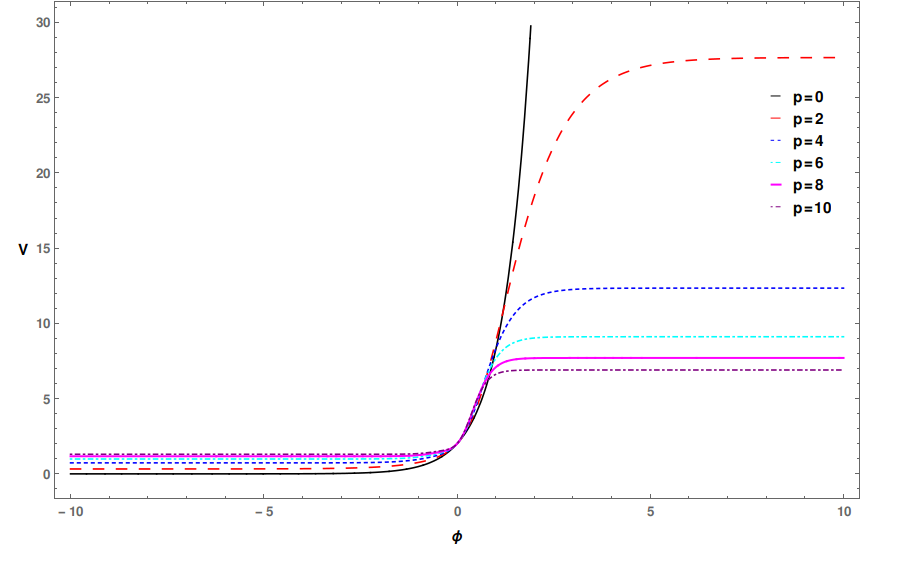}}
	\caption{\label{fig_potential}Variation of  potential involved in the units of ${\rm H_0}^2 {\rm M_P}^2$ with the scalar field for six different values of the parameter $p$.}
\end{figure}
Here again in \fig{fig_potential} we see that apart from $p=0$, the potential is very flat  during inflation and drop sharply near the end of inflation at $\phi=0$.

The amount of inflation is represented by the number of e-foldings, N, which can be exactly determined from \eq{eos_def} and we find that 
\bea
{\rm N}&=&\frac{1}{\sqrt{2}} \sinh (\phi{\rm M_P}^{-1})\  _2F_1\left(\frac{1}{2},\frac{1}{2}-\frac{p}{4};\frac{3}{2};-\sinh ^2(\phi{\rm M_P}^{-1})\right)
%&=&\frac{\phi }{\sqrt{2}}\left(1+\frac{p }{12}\phi ^2+\frac{p (3 p-4) }{480}\phi ^4+\frac{p (15 (p-4) p+64)}{40320} \phi ^6\right)+O\left(\phi^8\right)
\eea
\begin{figure}%[htb]
	\centerline{\includegraphics[width=16.cm, height=8.cm]{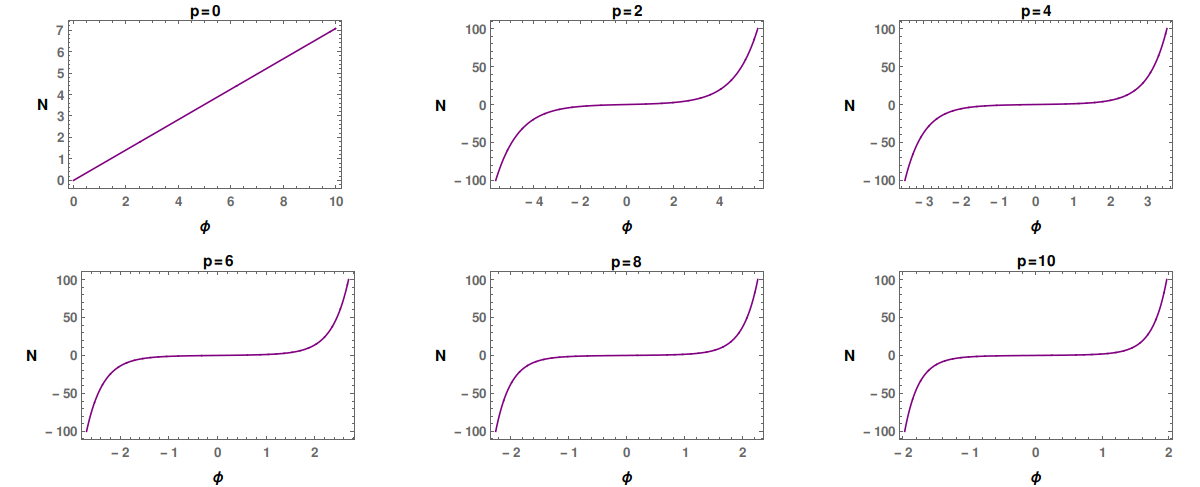}}
	\caption{\label{fig_efolds}Variation of e-foldings with the scalar field  for six different values of the model parameter $p$.}
\end{figure}
In order to have satisfactory answer to the Big-Bang puzzles generally 50-70 e-foldings are required. In \fig{fig_efolds} we have shown variation of N with the scalar field for different values of the model parameter. From the figure it is obvious that sufficient amount of inflation is achievable for any positive value of the parameter.  
%The inflationary observables are evaluated when there are still 55-65 e-foldings left before the end of inflation and within that window we can approximate N as 
%\beq
%{\rm N}\approx \frac{1}{\sqrt{2}}\sech^{-\frac{p}{2}}(\phi{\rm M_P}^{-1})
%\eeq

\begin{figure}%[htb]
	\centerline{\includegraphics[width=16.cm, height=10.cm]{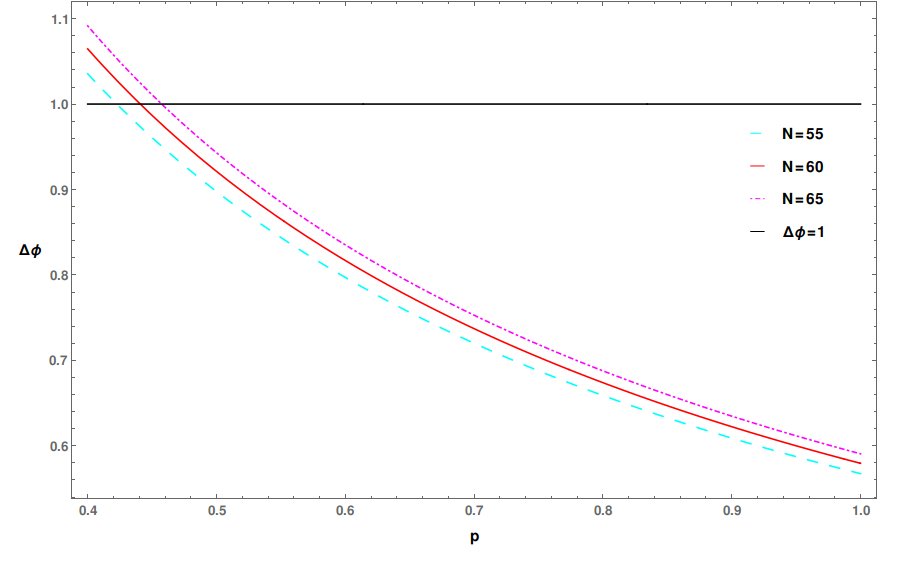}}
	\caption{\label{fig_deltaphi} Excursion of the inflaton during observable inflation taking $N=55, 60,65$ has been plotted with the model parameter $p$. The horizontal line corresponds to $\Delta\phi=1.0\ {\rm m_P}$.}
\end{figure}

Literally inflationary models are  categorized into two broad classes depending on the excursion of  inflaton, $\Delta\phi$, during observable inflation. One of them is small field inflation where $\Delta\phi<m_P$ and the large field modes where $\Delta\phi\geq m_P$. The energy scale of inflation is directly related to excursion of the inflaton.  Accordingly, tensor-to-scalar ratio determines variation of the scalar field during observable inflation, as shown in Ref.\cite{lyth1997}, in this present context we have found
\bea\label{lythd}
%\Delta\phi&=&\frac{m_P}{8\sqrt{\pi}}\int_0^{N_{\rm CMB}}\sqrt{r}\ dN\nonumber\\
%&=&\frac{m_P}{2\sqrt{2\pi}}\ \alpha^{-1}{\rm M_P}^{-1}\sinh^{-1}\left(\sqrt{2}\ \alpha  M_P \ N\right).
\Delta\phi&=&\frac{m_P}{8\sqrt{\pi}}\left(\phi_{\rm CMB}-\phi_{\rm  end}\right)
\eea
where $\rm m_P={2\sqrt{2\pi}}M_P$ is the actual Planck mass, $\phi_{\rm CMB}$ is the value of inflaton when the CMB scale left the horizon and $\phi_{\rm  end}$ is the value of inflaton at the end of inflation which is exactly zero for the model under consideration. For the large field models higher energy scale is required to match the current observations and as a result we get higher tensor-to-scalar ratio. In \fig{fig_deltaphi} we have shown variation of inflaton excursion in the unit of Planck mass $m_P$ with the model parameter. From the figure we see that there is a  small window where $\Delta\phi\geq  {\rm m_P}$ and for the rest of values of $p$, $\Delta\phi< {\rm m_P}$. This allows us to address both the large and small  scalar field models of  within a single framework.

Now in order to confront this model with recent observations we shall adopt slow-roll approximation which demands knowledge of another parameter given by
\bea
\eta_{_{\rm H}}&=& \frac{1}{2}\text{sech}^p(\phi{\rm M_P}^{-1}) \left(2-\sqrt{2} p \ \tanh (\phi{\rm M_P}^{-1}) \cosh ^{\frac{p}{2}}(\phi{\rm M_P}^{-1})\right)\nonumber\\
&=&\epsilon_{_{\rm H}}\left(1-\frac{1}{\sqrt{2}}p \ \epsilon_{_{\rm H}}^{-\frac{1}{2}}\left(1-\epsilon_{_{\rm H}}^{\frac{2}{p}}\right)^{\frac{1}{2}}\right)
\eea
\begin{figure}%[htb]
	\centerline{\includegraphics[width=16.cm, height=9.cm]{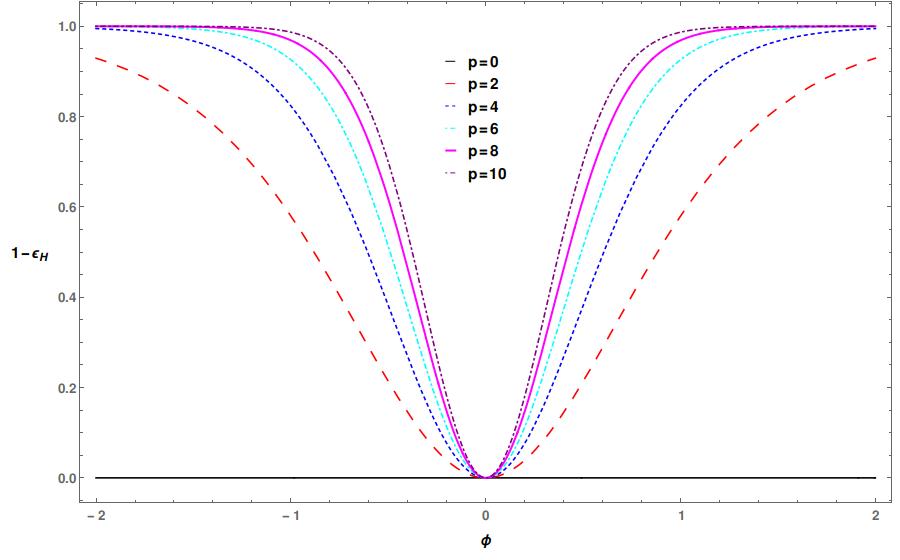}}
	\caption{\label{fig_epsilon} Variation of first slow-roll parameter with the scalar field for six different values of the model parameter, p.}
\end{figure}
\begin{figure}%[htb]
	\centerline{\includegraphics[width=16.cm, height=9.cm]{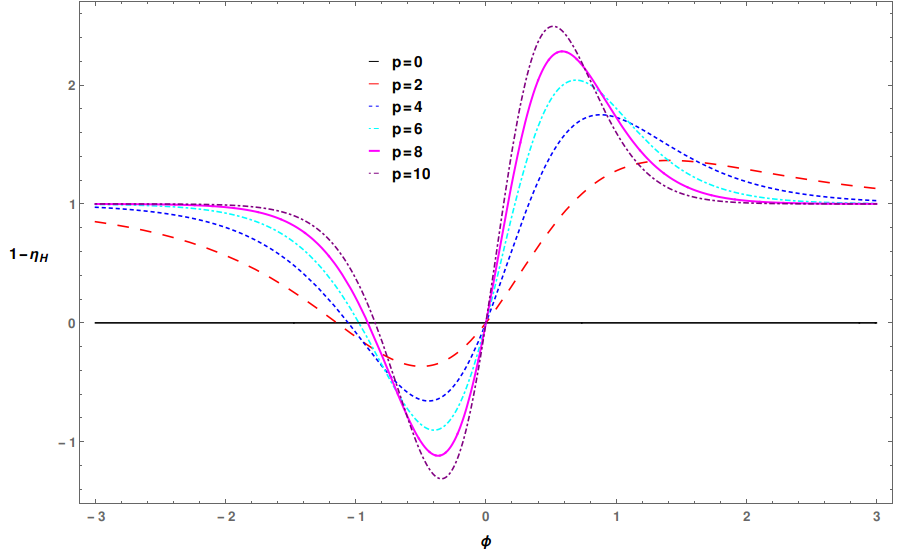}}
	\caption{\label{fig_eta} Variation of second slow-roll parameter with the scalar field for six different values of the model parameter.}
\end{figure}
In \fig{fig_epsilon} and \fig{fig_eta} we have shown the variation of slow-roll parameters. From the figures it is obvious that the model has slow-roll region almost for every  value of the model parameter, except for $p=0$. 

The most fascinating aspect of cosmological inflation is its ability to produce quantum mechanical seeds for cosmological perturbations observed in the large scale structure of the universe. Spectral tilt which measures the deviation from scale independence of the scalar curvature perturbation has been measured with great accuracy. Within the context of slow-roll inflation the first order expression for the spectral index is given by
 \bea
 n_{_S} &\approx&1 -4\epsilon_{_{\rm H}} + 2\eta_{_{\rm H}}\nonumber\\
 &=& 1-2\epsilon_{_{\rm H}}-{\sqrt{2}}p \ \epsilon_{_{\rm H}}^{\frac{1}{2}}\left(1-\epsilon_{_{\rm H}}^{\frac{2}{p}}\right)^{\frac{1}{2}}.
 \eea
\begin{figure}%[htb]
	\centerline{\includegraphics[width=16.cm, height=10.cm]{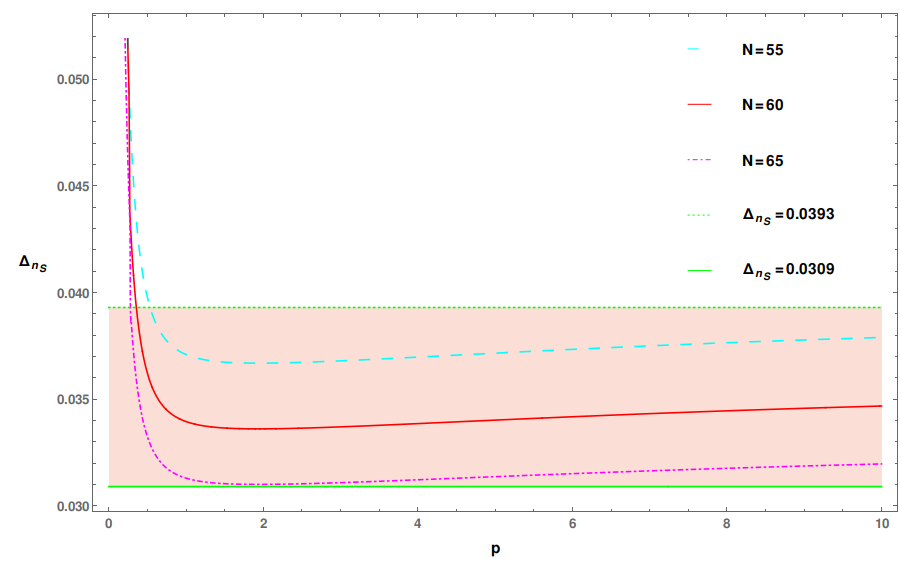}}
	\caption{\label{fig_spectral} The variation of spectral tilt with the model parameter $p$ for three different values of  ${\rm N}_{\rm CMB}$. The shaded region corresponds to current $1$-$\sigma$ bound from Planck 2018.}
\end{figure}
In \fig{fig_spectral} we have plotted the variation of spectral tilt, $\Delta_{n_{_S}}\equiv 1-n_{_S}$, with the model parameter along with the current $1$-$\sigma$ bound on the spectral tilt from Planck 2018 \cite{akrami2020planck}. We see that apart from very small values of p, the model is consistent with current data. This also allows us to set a lower bound on the model parameter $p\geq0.535,  0.358, 0.288$ for ${\rm N}_{\rm CMB}=55, 60,65$ respectively.

\begin{figure}%[htb]
	\centerline{\includegraphics[width=16.cm, height=9.cm]{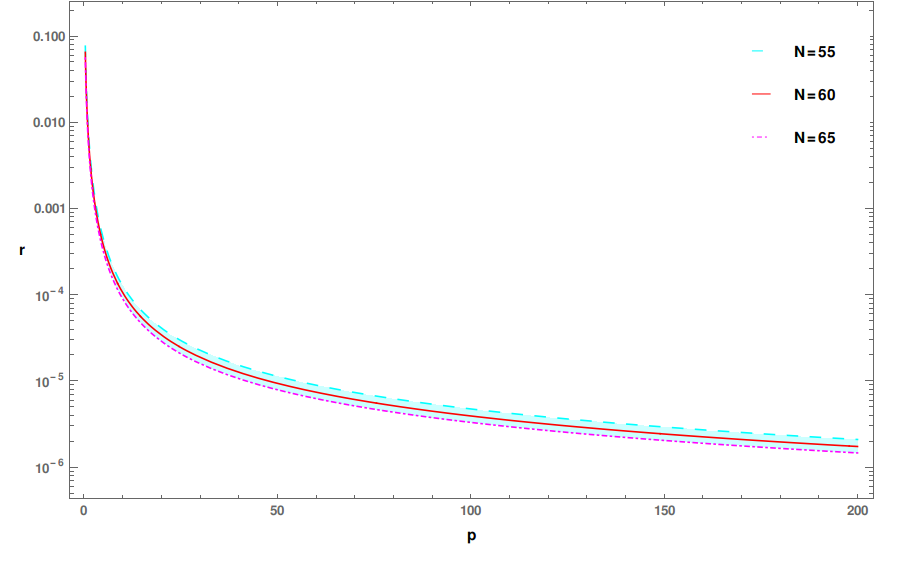}}
	\caption{\label{fig_tensor} The variation of tensor-to-scalar ratio  with the model parameter $p$ for three different values of ${\rm N}_{\rm CMB}$.}
\end{figure}

\begin{figure}%[htb]
	\centerline{\includegraphics[width=16.cm, height=9.cm]{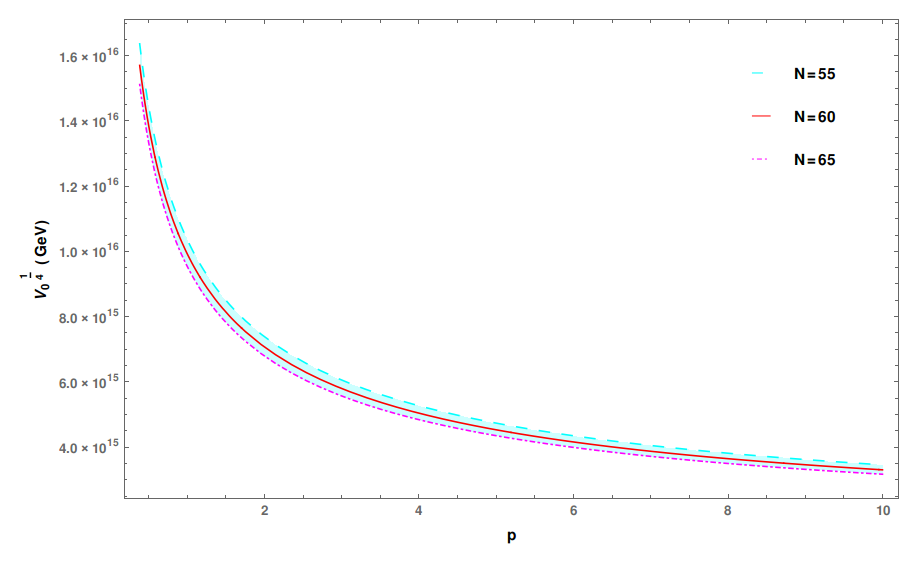}}
	\caption{\label{fig_hubble0} The variation of inflationary energy scale (in GeV) with the model parameter for three different values of ${\rm N}_{\rm CMB}$.}
\end{figure}

\begin{figure}%[htb]
	\centerline{\includegraphics[width=16.cm, height=10.cm]{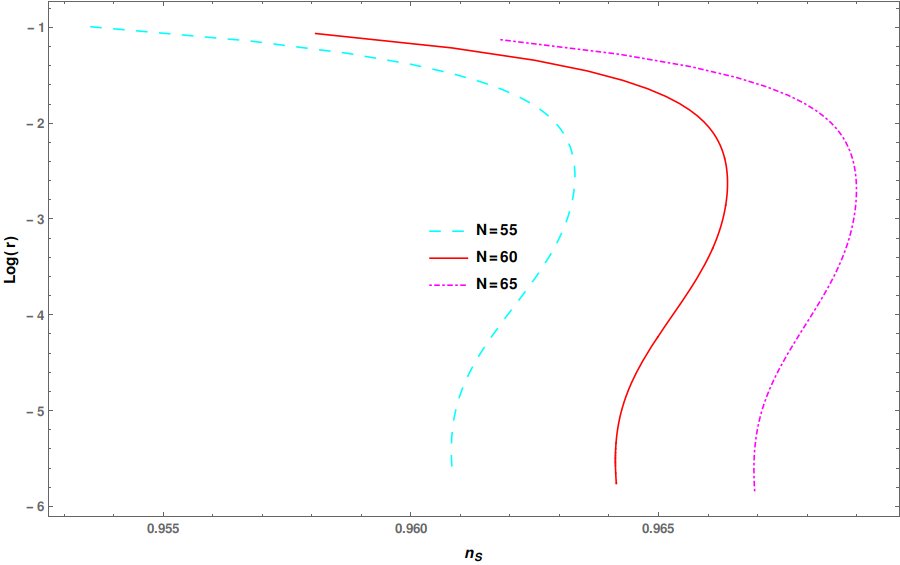}}
	\caption{\label{fig_rns} The logarithmic variation of tensor-to-scalar ratio with the scalar spectral index has been plotted for three different values of ${\rm N}_{\rm CMB}$.}
\end{figure}

Another exciting feature of cosmological inflation is the production of primordial gravity waves through tensor perturbation, the tensor- to-scalar ratio up to  first order in slow-roll parameters  turns out to be 
\bea
r&\approx&16\epsilon_{_{\rm H}}.
\eea
From \fig{fig_tensor} we recognize that as $p$ increases tensor-to-scalar ratio decreases. The resent analysis from Planck  suggests that $r<0.064$ \cite{akrami2020planck}, which further tightens the lower limit on the model parameter depending on ${\rm N}_{\rm CMB}$ and we find  that $p\geq0.385,  0.352, 0.325$ for ${\rm N}_{\rm CMB}=55, 60,65$ respectively.
The upper bound on tensor-to-scalar ratio  allows us to set the maximum value of the inflationary energy when the pivot scale exit the Hubble radius,  ${\rm V}_0^{1/4}<1.7\times 10^{16}\ { \rm GeV}$. In \fig{fig_hubble0} we have shown the variation  of associated energy scale  with the model parameter $p$ and the energy scale decreases as $p$ increases.
In \fig{fig_rns} we have plotted the logarithmic variation of tensor-to-scalar ratio with the spectral index. 

\begin{figure}%[htb]
	\centerline{\includegraphics[width=16.cm, height=9.cm]{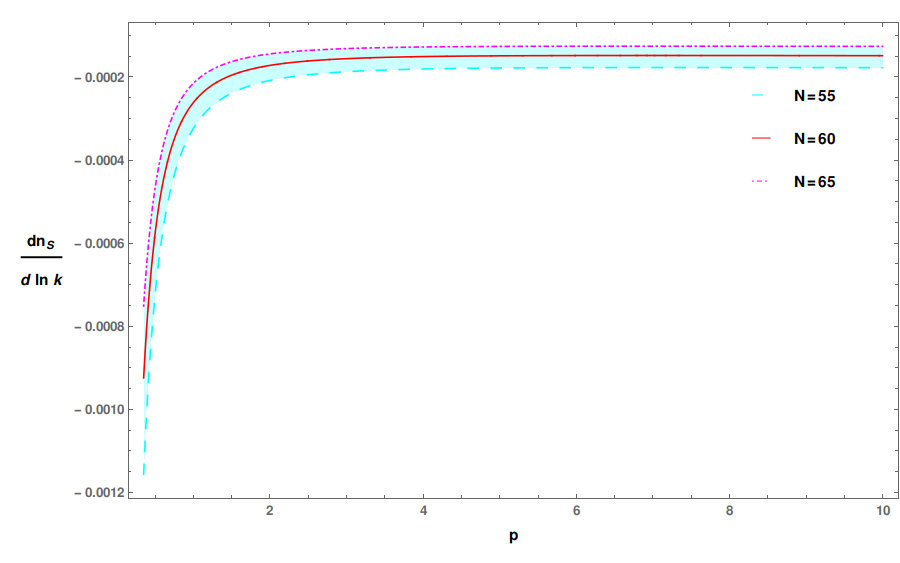}}
	\caption{\label{fig_running} Variation of the scalar running with the model parameter for three different values of ${\rm N}_{\rm CMB}$.}
\end{figure}

The scale dependence of the spectral index is measured by the scalar running. The Planck 2018 analysis has set $dn_{_S}/d\ ln\ k=-0.005\pm0.013$ when the running of the running is ignored. So scalar spectral index is almost scale invariant and we have found a very small negative running $\mathcal{O}(10^{-4})$ for most of the model parameter range as shown in \fig{fig_running}.
%%%%%%%%%%%%%%%%%%%%%%%%%%%%%%%%%%%%%%%%%%%%%%%%%%%%%%%%%%%%%%%%%%%%%%%%%%%%%%%%%%%%%%%%%%%%%%%%%%%%%%%%%%%%%%%%%%%%%%%%%%%%%%%%%%%%%%%%%%%%%%%%%%%%%%%%%%%%%%%%%%%%%%%%%%%%%%%%%%%%%%%%%%
\section{A Special Case: $p=2$}\label{p=2}
In this section we shall present results for a particular value of the model parameter $p=2$. Before producing expressions for the observable quantities, we first rewrite  the number of e-foldings in terms of  inflaton,
\beq\label{efoldings}
{\rm N}=\frac{1}{\sqrt{2}}\ \sinh\left(\phi{\rm M_{P}^{-1}}\right).
\eeq 
As a result the EoS can be rewritten as 
\beq\label{neos}
1+\omega=\frac{2}{3}\ \frac{1}{1+2{\rm N}^2}.
\eeq
Corresponding Hubble parameter now can be exactly determined using the EoS parametrization in the Hamilton-Jacobi formalism. In the present context we have found that 
\bea
{\rm H}&=&{\rm H}_0\exp\left[\frac{1}{\sqrt{2}}\tan^{-1}\left(\sinh(\phi{\rm M_{P}^{-1}})\right)\right]\nonumber\\
&=&{\rm H}_0\exp\left[\frac{1}{\sqrt{2}}\tan^{-1}\left({\sqrt{2}\rm N}\right)\right].
\eea
The associated inflaton potential 
\bea
{\rm V}&=&{\rm H_0}^2 {\rm M_P}^2 \left[3-\text{sech}^2(\phi{\rm M_P}^{-1} )\right] e^{\sqrt{2}\tan^{-1} \left(\sinh (\phi{\rm M_P}^{-1})\right)}\nonumber\\
&=& {\rm H_0}^2 {\rm M_P}^2 \left[3-\frac{1}{1+2{\rm N}^2}\right] e^{\sqrt{2}\tan^{-1}\left({\sqrt{2}\rm N}\right)}.
\eea
\begin{figure}%[htb]
	\centerline{\includegraphics[width=16.cm, height=9.cm]{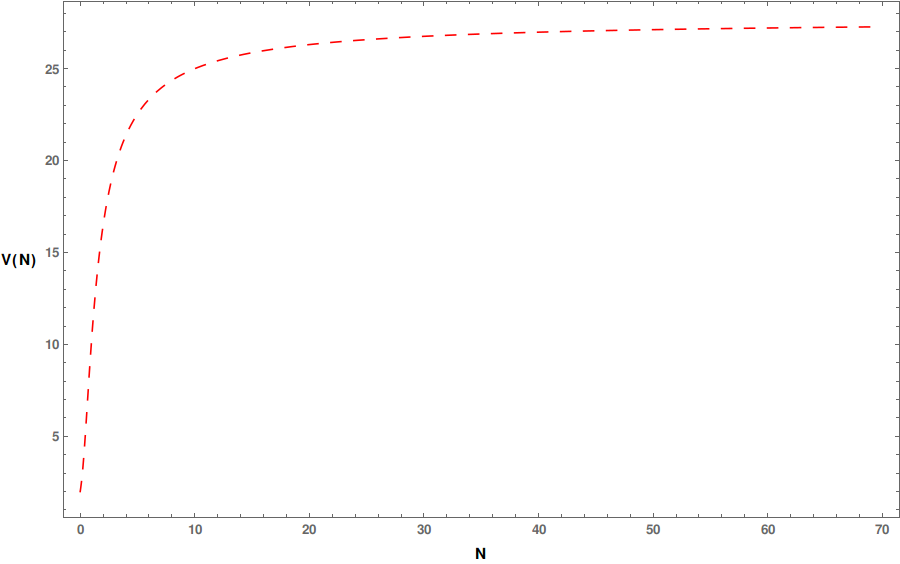}}
	\caption{\label{fig_potentialn} Variation of the inflaton potential  in the units ${\rm H_0}^2 {\rm M_P}^2$ with ${\rm N}$.}
\end{figure}

The slow-roll parameters in this special case turns out to be
\bea
\epsilon_{_{\rm H}}&=&\frac{1}{1
	+2{\rm N}^2},\quad \eta_{_{\rm H}}=\frac{1-2{\rm N}}{1+2{\rm N}^2}
\eea
The excursion of inflaton during observable inflation turn out to be 
\bea
\Delta\phi&=&\frac{\rm m_P}{8\sqrt{\pi}}\left(\phi-\phi_{\rm  end}\right)\nonumber\\
&=&\frac{\rm m_P}{8\sqrt{\pi}}\sinh^{-1}\left(\sqrt{2}{\rm N}\right).
\eea
From the above expression it is obvious that for $p=2$ inflaton variation is sub-Plankian and it corresponds to small field inflation.

The amplitude of curvature perturbations within the slow-roll approximation is given by 
\bea
P_{\cal R}&\approx&\frac{1}{8\pi^2 \rm M_{P}^2} \frac{{\rm H}^2}{\epsilon_{_{\rm H}}}\nonumber\\
&=&\frac{{\rm H}_0^2}{8\pi^2 \rm M_{P}^2}\quad(1+2{\rm N}^2)\exp\left[\sqrt{2}\tan^{-1}\left({\sqrt{2}\rm N}\right)\right]\label{sr_pr_hubble}\\
&=&\frac{{\rm V}_0}{8\pi^2 \rm M_{P}^4}\quad(1+2{\rm N}^2)\exp\left[\sqrt{2}\tan^{-1}\left({\sqrt{2}\rm N}\right)\right]\label{sr_pr}
\eea
where in the last line we have defined the inflationary energy scale ${\rm V}_0\equiv{\rm H}_0^2{ \rm M_{P}}^2$. The \eq{sr_pr_hubble} we can be inverted to get the  value of Hubble parameter during inflation 
\beq
{\rm H}_0\sim2\sqrt{2}\pi {\rm M_{P}}\ P_{\cal R}^{\frac{1}{2}}\left(1+2{\rm N}^2\right)^{-1/2}\exp\left[-\frac{1}{\sqrt{2}}\tan^{-1}\left({\sqrt{2}\rm N}\right)\right].
\eeq
Current PLanck 2018 upper bound on the Hubble parameter during inflation is ${\rm H}_0<2.7\times 10^{-5}\ { \rm M_{P}}$. For the model under consideration we have found that 
 ${\rm H}_0<1.75613\times 10^{-6}\ { \rm M_{P}}$, a order lower than present bound.  

The spectral index turns out to be
\bea
 n_{_S} &\approx&1-\frac{2+4\rm N}{1+2{\rm N^2}}\nonumber\\
 &\approx& 1-\frac{2}{\rm N}
 \eea
 The tensor-to-scalar ratio is now
 \bea
 r&\approx&\frac{16}{1+2{\rm N^2}}\nonumber\\
 &\approx&\frac{8}{{\rm N^2}}.
 \eea
 From the above outcomes it is clear that the particular case for $p=2$ very closely resembles $\alpha$-Attractor class of inflationary models with $\alpha=\frac{2}{3}$ which is special as it corresponds to unit moduli space curvature. 
 
\iffalse The spectral index and scalar tensor ratio
 \beq
1- n_{_S} \approx\frac{r}{8}\left(1+2{\rm N}\right)
 \eeq
\bea
\alpha_{_S}&\approx&-\frac{4}{\left(1+2\rm N^2\right)^2} \left(2\rm N^2+2\rm N-1\right)\nonumber\\
&\approx&-\frac{2}{\rm N^2} 
\eea
\fi

%%%%%%%%%%%%%%
%%%%%%%%%%%%%%%%%%%%%%%%%%%%%%%%%%%%%%%%%%%%%%%%%%%%%%%%%%%%%%%%%%%%%%%%%%%%%%%%%%%%%%%%%%%%%%%%%%%%
\section{Conclusion}\label{conclusion}
Inflation remains the best tool to explain early universe scenario, in fact it is the only mechanism that is compatible with recent observations when combined with the Big-Bang. Precise data of late has eliminated many inflationary models but still allowing  plenty of them. As a consequence more precision is required towards a specific compelling model of inflation. 

In this present analysis  we have found that a simple single parameter EoS has quite remarkable fit to the latest  data. Not only that, the prediction from this particular inflationary EoS can efficiently explain wide range of tensor-to-scalar ratio which many existing inflationary models fails to address. The EoS for a specific value of the model parameter mimics the $\alpha$-Attractor class of inflationary models with $\alpha=\frac{2}{3}$. The present and future observational probes are targeting the primordial gravity waves, whose detection would certainly help to discriminate between inflationary models. The tensor-to-scalar ratio being the most powerful up-to-date  eliminator and the  model discussed here is capable of producing almost  any value of $r$, it might have the endurance and vision to emerge as the winner in long term. 

\bibliographystyle{unsrt}
\bibliography{references}

\end{document}